\documentclass[prl,amsmath,amssymb,twocolumn, showpacs, superscriptaddress,10pt]{revtex4}

\usepackage[english]{babel}
\usepackage[utf8]{inputenc}
\usepackage[T1]{fontenc}
\usepackage{amsmath}
\usepackage{hyperref}
\usepackage{graphicx}
\usepackage{amsfonts}
\usepackage{amsthm}

\theoremstyle{remark}

\usepackage{color}
\definecolor{Blue}{rgb}{0.00, 0.00, 1.00}
\definecolor{Red}{rgb}{1.00, 0.00, 0.00}

\newcommand{\be}{\begin{equation}}
\newcommand{\ee}{\end{equation}}
\newcommand{\bea}{\begin{eqnarray}}
\newcommand{\eea}{\end{eqnarray}}

\newcommand{\beq}{\begin{equation}}
\newcommand{\eeq}{\end{equation}}
\newcommand{\beqn}{\begin{eqnarray}}
\newcommand{\eeqn}{\end{eqnarray}}

\def\Xint#1{\mathchoice
   {\XXint\displaystyle\textstyle{#1}}%
   {\XXint\textstyle\scriptstyle{#1}}%
   {\XXint\scriptstyle\scriptscriptstyle{#1}}%
   {\XXint\scriptscriptstyle\scriptscriptstyle{#1}}%
   \!\int}
\def\XXint#1#2#3{{\setbox0=\hbox{$#1{#2#3}{\int}$}
     \vcenter{\hbox{$#2#3$}}\kern-.5\wd0}}

\def\dashint{\Xint-}

\begin{document}

\setlength{\abovedisplayskip}{5pt}
\setlength{\belowdisplayskip}{5pt}

\title{Manifolds in high dimensional random landscape: complexity of stationary points
and depinning}

\author{Yan V. Fyodorov}
\affiliation{King's College London, Department of Mathematics, London  WC2R 2LS, United Kingdom}
\author{Pierre Le Doussal}
\affiliation{Laboratoire de Physique Th\'eorique de l'Ecole Normale Sup\'erieure,
PSL Research University, CNRS, Sorbonne Universit\'es,
24 rue Lhomond, 75231 Paris, France}

\begin{abstract}
We obtain explicit expressions for the annealed complexities associated respectively with the total number of (i) stationary points and (ii) local minima of the energy landscape for an elastic
manifold with internal dimension $d<4$ embedded in a random medium of dimension $N \gg 1$ and confined
by a parabolic potential with the  curvature parameter $\mu$. These complexities
are found to both vanish at the critical value $\mu_c$ identified as the Larkin mass.
For $\mu<\mu_c$ the system is in complex phase
corresponding to the replica symmetry breaking in its $T=0$ thermodynamics.
The complexities vanish respectively quadratically (stationary points)
and cubically (minima) at $\mu_c^-$. For $d\geq 1$ they admit  a finite ''massless'' limit
$\mu=0$ which is used to provide an upper bound for the depinning threshold
under an applied force.

\end{abstract}



\maketitle

Numerous physical systems can be modeled by a collection of points or particles coupled by
an elastic energy, usually called an elastic manifold, submitted to a random potential
(see  \cite{TGPLDBragg2} for a review). They are often called "disordered elastic systems" and generically exhibit pinning in their statics
and depinning transitions and avalanches in their driven dynamics
\cite{Fis85,TGPLDBragg1a,RosKra02,LeDWie04,LeDWie13}.

The manifold can be parameterized by a $N$-component field ${\bf u}(x) \in \mathbb{R}^N$,
where $x$ spans an internal space $x \in \Omega$, either $L^d$ points
on a discrete lattice $\Omega \subset \mathbb{Z}^d$,
or $\Omega \subset  \mathbb{R}^d$ of volume $L^d$ in the continuum setting.
The energy functional \cite{MezPar91,MP2}
\begin{equation}\label{landscape}
{\cal H}[{\bf u}]=\sum_{x,y} {\bf u}(x) \cdot (\mu_0 \mathbf{1} - t_0 \Delta)_{xy} \cdot {\bf u}(y)
+ \sum_{x} V(\mathbf{u}(x),x)
\end{equation}
is the sum of an elastic energy, given by the (discrete) Laplacian matrix
$- t_0 \Delta_{xy}$, $t_0>0$, a quadratic confining
energy controlled by the curvature  parameter  $\mu_0>0$ (or, alternatively, the ''mass'' $m=\sqrt{\mu_0}$) and
a centered Gaussian random potential with covariance
\begin{equation}\label{cov}
\overline{ V(\mathbf{u}_1,x_1) V(\mathbf{u}_2,x_2) } = N\: B\left(\frac{(\mathbf{u}_1-\mathbf{u}_2)^2}{N}\right) \delta_{x_1,x_2}
\end{equation}
parametrized by a function $B(z)$. This random potential is thus
uncorrelated in the internal space and statistically translational invariant
in the embedding space. We use periodic boundary conditions,
i.e. the Laplacian eigenmodes are plane waves $\sim e^{i k x}$
with eigenvalues $\Delta(k)$. Examples are $\Delta(k)= 2 (\cos k-1)$
with $k=2 \pi n/L$, $n=0,..L-1$ in $d=1$, and for the continuum
model  $\Delta(k)= -k^2$ in any dimension with $k \in \mathbb{R}^d$.

The energy landscape provided by the functional (\ref{landscape})  is complex and necessarily high-dimensional, i.e. involves many interacting and competing degrees of freedom, leading to glassy behavior.
This necessitates to use methods and ideas of statistical mechanics of disordered systems such as Replica Symmetry Breaking (RSB),
Functional Renormalization Group, etc. for understanding their properties \cite{MP2,BalBouMez96,PLDKWLargeNDetails,LDMW}.
The problem of characterizing random high-dimensional landscapes by understanding the statistics 
of their stationary points (minima, maxima and saddles),
defined in our model by the condition $\frac{\delta {\cal H}}{\delta {\bf u}(x)}=0$ for all $x$, has attracted considerable recent interest 
in pure and applied mathematics, see e.g. \cite{AufBenCer13,AufBen13,SubZei15,RBBC2018,Sub2017},   as well as in theoretical physics, see \cite{AnnCavGiaPar03,Fyo04,DouShiZel04,Par05,BraDea07,FyoWil07,FyoNad12,EasGutMas16} and references to earlier works in \cite{Fyo15}. In that context the nontrivial, glassy physics is closely
associated with the exponential growth of the number of stationary points of various indices (number of unstable directions), with the corresponding rates of growth being known as the {\it landscape complexities}, see  (\ref{defSigma}) below. Vanishing of those complexities as a function of parameters then signals of a phase transition, which for an elastic manifold is naturally associated with depinning manifesting itself as a {\it topology trivialization} phenomenon \cite{FLRTa}.

For the energy functional \eqref{landscape}-\eqref{cov} in the simplest 'toy model' limit
 $d=0$ with no elastic interactions (when $x$ is essentially a single point, $L^d=1$), the mean number of stationary points, $\overline{ {\cal N}_{\rm tot}}$, and of stable equilibria (local minima), $\overline{ {\cal N}_{\rm st}}$, of the
landscape were investigated in the limit of large $N \gg 1$ in \cite{Fyo04,FyoWil07,FyoNad12}. It was found that a sharp transition occurs from a 'simple' landscape for $\mu_0>\mu_{c}$  with typically only a single stationary point (the minimum) to a complex ('glassy') landscapes for $\mu_0<\mu_c$ with exponentially many stationary points. Such transition has been shown to coincide with the onset of RSB  in the associated
model of statistical mechanics \cite{FyoWil07}, see also related studies in\cite{AufBenCer13,AufBen13,SubZei15,RBBC2018}.
From the other end, the case of an elastic string $d=1$ in dimension $N=1$ has been recently addressed in \cite{FLRTa},
where relations with disordered Schroedinger operators and Anderson localization problems have been revealed and exploited.

The goal of the present work is to demonstrate that as long as $N\to \infty$, the problem of characterizing the corresponding annealed complexities defined as
\be \label{defSigma}
\Sigma = \lim_{N \to \infty} \frac{\log \overline{{\cal N}_{\rm tot} }}{N L^d}
\quad , \quad \Sigma_{\rm st}  = \lim_{N \to \infty} \frac{\log \overline{{\cal N}_{\rm st} }}{N L^d}
\ee
can be as completely investigated for disordered elastic manifolds of internal dimensions $1 \leq d < 4$ as for the limiting 'toy model' case $L^d=1$.
In doing this we combine the approaches of \cite{FLRTa} with the insights from our earlier study of Hessians for high-dimensional manifolds \cite{UsHessToy,UsHess}. To connect to the phenomenon of depinning,
we also calculate the complexity in presence of an applied force along the direction $i=1$, which
corresponds to the change in the model \eqref{landscape}
\be \label{force1}
{\cal H}[u] \to {\cal H}[u] - \sqrt{N} \sum_x f(x) u_1(x)
\ee
Depinning is usually discussed for a uniform force $f(x)=f$ in the context of $T=0$ relaxational
dynamics. Here we define the depinning threshold as the minimal force $f_c$ such that
there exist no set of generic initial conditions leading to pinning
\cite{footnote4}. Such a definition implies that $f_c$ is the force at which
all stable equilibria in a {\it typical} landscape disappear.
Hence it is the value of $f$ where the quenched complexity $\lim_{N \to \infty}  \frac{\overline{\log {\cal N}_{\rm st}}}{N L^d}$ vanishes. Convexity of the logarithm implies the bound for the depinning threshold
\be \label{bound}
f_c \leq f_c^{\rm st}
\ee
where $f_c^{\rm st}$ is obtained below in \eqref{fcst}. This generalizes a similar bound obtained in the
case $N=1$ in \cite{FLRTa}.

Our starting point is the explicit form for the stationarity condition, i.e. mechanical
equilibria, for the energy functional ${\cal H}[{\bf u}]$, as
\be \label{Hessian0}
\frac{\partial {\cal H}[{\bf u}] }{ \partial {\bf u}(x)} = \sum_y (\mu_0 \mathbf{1} - t_0 \Delta)_{xy} \cdot {\bf u}(y) +
\frac{\partial}{\partial {\bf u}(x)}V({\bf u}(x),x) = 0
\ee
The total number of solutions of these equations in each realization of the random potential is given by the Kac-Rice type formula
\bea \label{ntot}
{\cal N}_{\rm tot} = \int \, {\cal D}{\bf u}(x) \prod_{x} \delta^N\left( \frac{\partial {\cal H}[{\bf u}] }{ \partial {\bf u}(x)}\right)
\, |\det {\cal K}^0[{\bf u}]|
\eea
where ${\cal K}^0[{\bf u}]$ is the
$N L^d \times N L^d$ Hessian matrix around the configuration ${\bf u}(x)$
\bea \label{Hessian}
{\cal K}^0_{ix,jy}[{\bf u}] &=& \frac{\partial ^2}{\partial u_{i}(x) \partial u_j(y)}{\cal H}[{\bf u}] \\
&=&\delta_{ij} (\mu_0 \mathbf{1} - t_0 \Delta)_{xy} + \delta_{xy} \frac{\partial ^2}{\partial u_{i}\partial u_j}V({\bf u}(x),x)\,.
\nonumber
\eea
We will also be interested in the number of local minima ${\cal N}_{\rm st}$, obtained
from \eqref{ntot} by inserting the step-function factor $\theta({\cal K}^0[{\bf u}])$ which selects
only positive definite Hessians. The simplest yet informative quantities are the
mean values $\overline{{\cal N}_{\rm tot} }$ and $\overline{{\cal N}_{\rm st} }$. From a simple generalization of
the considerations in \cite{FLRTa} those values can be represented as 
\be \label{Nmoy}
\overline{{\cal N}_{\rm tot} }= \frac{ \overline{|\det {\cal K}[{\bf 0}]|}}{[\det(\mu - t \Delta)]^N}
\quad , \quad \overline{{\cal N}_{\rm st} }= \frac{ \overline{|\det {\cal K}[{\bf 0}]| \theta({\cal K}[{\bf 0}])}}{[\det(\mu - t \Delta)]^N}
\ee
where we have defined $\mu=\mu_0/J$, $t=t_0/J$ and ${\cal K} = {\cal K}^0/J$, scaled with the disorder strength parameter
$J^2 = 4 B''(0)$.
From these we define the annealed complexities as shown in \eqref{defSigma}.

The Hessian matrix has a block structure \cite{UsHess}, with blocks of size $N \times N$. Only
diagonal blocks contain random Gaussian entries. Different blocks are coupled
by the lattice Laplacian. From \eqref{cov} follows the equality in distribution
at a given $x$
\be
\frac{\partial ^2 V({\bf u},x)}{\partial u_{i}\partial u_j}|_{{\bf u}=0}
\equiv
J \left[\xi(x) \delta_{ij}  + H(x)_{ij} \right]\,,
\ee
where $H(x)$ are a set of $L^d$  GOE(N) matrices independent for different $x$ ( each distributed with $P(H) \sim e^{- \frac{N}{4} ({\rm Tr} H)^2}$ implying asymptotically the spectrum supported in $[-2,2]$) and $\sqrt{N} \xi(x)$ are i.i.d. standard Gaussian variables independent of the $H(x)$.
This naturally leads to the decomposition ${\cal K} = K + X + \mu I$,
where $X_{ix,jy}=  \xi(x) \delta_{ij} \delta_{xy}$
 and $K_{ix,jy}= H_{ij} \delta_{xy} - t \delta_{ij} \Delta_{xy}$, e.g. for $d=1$, $L=2$
\be
\!\! \! X =  \begin{pmatrix} \xi(1) I_N & 0  \\
0 & \xi(2) I_N
\end{pmatrix}
, K = \begin{pmatrix}
 H(1) + 2  t I_N & -  2 t I_N  \\
- 2 t  I_N &  H(2) + 2  t I_N
\end{pmatrix}
\ee

{\bf Total complexity}. Let us now evaluate the numerator in \eqref{Nmoy}
as
\be \label{start}
\!\! \overline{|\det {\cal K}[{\bf 0}]|} = \prod_x \int_{\mathbb{R}}  \frac{d\xi(x) e^{- N \frac{\xi(x)^2}{2}}}{\sqrt{2 \pi/N}}
\langle | \det ( K + X + \mu I ) | \rangle_{\rm GOE's}
\ee
where $\langle \dots \rangle_{\rm GOE's}$ denotes averaging over all $L^d$
independent GOE matrices. The exact calculation of the r.h.s. of \eqref{start}
is challenging due to the modulus of the determinant, and has been performed
only in the toy model case $L^d=1$ \cite{Fyo04,FyoWil07,FyoNad12}. Here we conjecture that,
to leading order for $N \to +\infty$ at fixed $L\ge 1,0\le d\le 4$ one is allowed to replace
\be
  \left\langle | \det (K + X + \mu I)  | \right\rangle_{\rm GOE's} \approx
 e^{\,\left\langle {\rm Tr}  \log | K + X +  \mu I |  \right\rangle_{\rm GOE's}}
\ee
For $L^d=1$ it is in fact a theorem following from large deviations,
and is a consequence of the rigidity
of the spectrum of GOE matrices, see e.g. discussion in \cite{AufBenCer13,AufBen13}. We expect that such a rigidity
should also hold in the considered limit of block banded models. { Indeed, as has been shown rigorously for a slightly different, though closely related
model of random block band matrices \cite{Shcher14}, setting the block size $N$ to infinity for any finite number of blocks ensures that the eigenvalue correlations remain of Wigner-Dyson type implying
 spectral rigidity. Our conjecture then amounts to assuming that sending the number of blocks to infinity after $N\to \infty$ is not destroying the rigidity, which looks as a very plausible assumption. The aim of our paper is to explore the consequences of this conjecture, which, as we shall see are very interesting. In particular, we show in \cite{SM} that the expression for the mean resolvent is correctly reproduced (see below), providing a certain consistency check to our assumptions}.
This leads to
\be \label{start2}
\overline{|\det {\cal K}[{\bf 0}]|}|_{N \gg 1} \sim \prod_x \int_{\mathbb{R}}
 \frac{d\xi(x)}{\sqrt{2 \pi/N}}  e^{- N S[\xi] }
\ee
with
\be \label{Sx}
S[\xi]= \sum_x \frac{1}{2} \xi(x)^2 - \frac{1}{N} \left\langle {\rm Tr}  \log | K + X +  \mu I |  \right\rangle_{\rm GOE's}
\ee
The integral is dominated at large $N$ by the saddle point for $\xi(x)$ given by
\be
\xi(x) = \frac{1}{N}  \left\langle {\rm Tr}  (K + X +  \mu I)^{-1}_{xx}  \right\rangle_{\rm GOE's}
\ee
This equation has a solution independent of $x$, $\xi(x)=\xi_*$, where $\xi_*$ solves the
equation
\be \label{sp0}
\xi_* = f'(\xi_* + \mu)  \quad , \quad  f(\xi) := \int d\lambda \ln|\lambda+\xi| \, \rho_{K}(\lambda)
\ee
where $\rho_K(\lambda)$ is the mean eigenvalue density of the random matrix $K$
in the limit $N \to +\infty$ which was studied by us recently\cite{UsHess}.
It is given by the imaginary part of the resolvent $i r_\lambda$
\be \label{dens}
\rho_{K}(\lambda) = \frac{1}{\pi} {\rm Im} (i r_\lambda)|_{{\rm Im} \lambda=0^-}~,~
i r_\lambda := \frac{1}{N L^d} \langle {\rm Tr} (\lambda - K)^{-1} \rangle
\ee
which satisfies the following self-consistent equation\cite{UsHessToy,KhorPast1993}
\be \label{sc}
i r_\lambda = \int_k \frac{1}{\lambda \,  + t \Delta(k)   - i r_\lambda}
\ee
where we denote $\int_k = \frac{1}{L^d} \sum_k  \equiv \int \frac{d^d k}{(2 \pi)^d}$ so
that our formula are valid both for discrete and continuum models (in the latter case $\sum_x \equiv \int d^dx$).
The complexity defined in \eqref{defSigma} is obtained from the value at the saddle point as
\be \label{sig1}
\Sigma(\mu) =  - \frac{1}{2} (\xi^*)^2 +  f(\xi^* + \mu)  - \int_k  \ln (\mu - t \Delta(k))
\ee
Let us note that at the saddle point for $\xi$, we have
\bea \label{sp2}
\xi^* = - {\rm Re} [ i r_{- \xi^*-  \mu + i 0^+} ]
\eea
since by definition of $f(\xi)$ one has
\be
f'(\xi) = \dashint  \frac{d\lambda \rho_{K}(\lambda)}{\xi+\lambda}
= {\rm Re}\int \frac{d\lambda \rho_{K}(\lambda)}{\xi+\lambda- i 0^+}
=
- {\rm Re} ( i r_{- \xi + i 0^+} ) \nonumber
\ee
To evaluate the real part in the r.h.s. of \eqref{sp2}
we separate the real and imaginary parts
\bea
i r_\lambda = x_\lambda + i y_\lambda
\eea
and \eqref{sc} then leads to the equivalent pair of equations
\bea \label{solu2}
&& x_\lambda = \int_k \frac{ \lambda - x_\lambda
+  t \Delta(k)}{(\lambda - x_\lambda +  t \Delta(k))^2 + y_\lambda^2}  \\
&& y_\lambda = y_\lambda \int_k \frac{ 1}{(\lambda - x_\lambda +  t \Delta(k))^2 + y_\lambda^2}
\eea
where $y_\lambda \geq 0$.
Substituting $\lambda=- \xi_*-  \mu$ in these equations, one obtains
the pair of equations
\bea \label{solu}
&& \xi_* = \int_k \frac{ \mu -   t \Delta(k)}{( \mu -  t \Delta(k))^2 + y^2}  \\
&& y = \int_k \frac{y}{( \mu -  t \Delta(k))^2 + y^2}  \label{soluy}
\eea
since $\lambda-x_\lambda|_{\lambda=- \xi_*-  \mu}=-\mu$
using \eqref{sp2}. Here $y \geq 0$ is a variable which
should be eliminated between the two equations to obtain the value $\xi^*$
at the saddle point as a function of $\mu$. Obviously there are two
phases depending on whether $y=0$ or $y >0$.
Noting, from \eqref{dens}, that
\be \label{remark}
y=y_{\lambda}=y_{- \xi_*-\mu}= \pi \rho_K(- \xi_*-\mu)
\ee
we see that these two phases also correspond to $- \xi_*-\mu$
belonging or not to the support of the mean eigenvaue density of $K$.

{\it Simple phase.} In this phase $y=0$. We now show that the
complexity vanishes in this phase. One has from \eqref{solu}
\be \label{solu2}
 \xi_* = \int_k \frac{1}{\mu -  t \Delta(k)}
\ee
Taking a derivative w.r.t. $\mu$ in \eqref{sig1} and
using the saddle point condition \eqref{sp0}
we see that the derivative
\be \label{dersig}
\partial_\mu \Sigma(\mu) =
f'(\xi_* + \mu) - \int_k \frac{1}{\mu - t \Delta(k)} = 0
\ee
vanishes in view of \eqref{sp0} and \eqref{solu2}. Since $\Sigma(+\infty)=0$
we obtain that $\Sigma(\mu)$ is zero everywhere in this phase.

{\it Complex phase}. In this phase $y > 0$. From \eqref{soluy} the boundary of this phase
is given by $\mu= \mu_c$ which solves
\be \label{refs}
1 = \int_k \frac{1}{( \mu_c -  t \Delta(k))^2}
\ee
This value $\mu_c$  defines the so-called Larkin mass for our model (see discussion and references in \cite{UsHess}) and the criterion
\eqref{refs} is known to signal a continuous transition towards a RSB phase for $\mu < \mu_c$
in the corresponding
statistical mechanics model at $T=0$. In that phase metastability dominates thermodynamics.
It is consistent to find here that for $\mu< \mu_c$ the complexity
$\Sigma$ is non zero, as we now show.

Replacing $f'(\xi_* + \mu)$ with $\xi_*$ into \eqref{dersig} using \eqref{sp0},
and integrating over $\mu$ from $\mu_c$, we obtain
\be \label{SS}
 \Sigma(\mu)= - \int_{\mu}^{\mu_c} d \tilde \mu \left( \xi_*(\tilde \mu) -  \int_k (\tilde \mu - t \Delta(k))^{-1} \right)
\ee
Here we denote $(\xi_*(\mu),y(\mu) > 0)$ the solution of the system of equations \eqref{solu} for $\mu<\mu_c$.
Substituting $\xi_*(\mu)$ from \eqref{solu}, we obtain our final result for the complexity as
\be \label{SS2}
 \Sigma(\mu) = \int_{ \mu}^{ \mu_c} d\tilde \mu  \int_k \frac{y(\tilde \mu)^2}{(\tilde \mu - t \Delta(k)) \left( (\tilde \mu - t \Delta(k))^2 + y(\tilde \mu)^2\right)}
\ee
where $y(\mu)$ is determined by
\be \label{ymu}
1 = \int_k \frac{ 1}{(\mu -  t \Delta(k))^2 + y(\mu)^2}
\ee
When $\mu \to \mu_c^-$ we find that $y(\mu) \to 0$ and $\Sigma(\mu) \to \Sigma(\mu_c)=0$.
The transition between the phases is continuous. Performing the expansion for $\mu=\mu_c(1-\delta)$
for small $\delta>0$ we find \cite{SM}
\be \label{expansion0}
\Sigma(\mu) = \frac{I_3(\mu_c)^2}{I_4(\mu_c)} \mu_c^2 \delta^2 + O(\delta^3)
\ee
where we defined $I_p(\mu)=\int_k \frac{1}{( \mu -  t \Delta(k))^p}$.
There is thus a jump in the second derivative of the complexity at the transition.

Let us give a few examples. In $d=0$, i.e. $L=1$ one finds $y(\tilde \mu)^2 = 1- \tilde \mu^2$, hence
$\mu_c=1$ and $ \Sigma(\mu) = \frac{1}{2} (\mu^2 - 1) - \log \mu$ recovering the result of \cite{Fyo04}.
For the continuum model  $\Delta(k)=- k^2$ in the $L \to +\infty$ limit the expression for the Larkin mass has a very
explicit form for any $d<4$ \cite{SM}:
\be\label{Larkinmasscont}
\mu_c=\left(\frac{\tilde I_2}{t^{d/2}}\right)^{\frac{2}{4-d}}, \quad \tilde I_2=\frac{1}{2^d\pi^{d/2}}\Gamma\left(\frac{4-d}{2}\right)
\ee
so that
\[
\mu_c|_{d=1}=\left(\frac{1}{4\sqrt{t}}\right)^{2/3},  \quad \mu_c|_{d=3}=\left(\frac{1}{8\pi\,t^{3/2}}\right)^{2}
\]
and $\mu_c|_{d=2}=\frac{1}{4 \pi t}$. Moreover, the total complexity for $d=2$ is also very explicit \cite{SM}:
\be
\Sigma(\mu) =  \frac{ \mu}{8 \pi t} \left(4 \pi t \mu Z^2 -  \log(1 + Z^2)  \right) \quad , \quad
 \mu=\frac{\tan^{-1} Z}{4 \pi t\, Z}
\ee

The massless limit $\mu \to 0^+$ is of great interest since the system in $d>0$ becomes
critical with power law roughness of the ground state. As in \cite{FLRTa} for the case $d=1$, $N=1$, our results
can be expressed in terms of the Larkin length $L_c$
as $L_c \propto \Sigma(0)^{-1/d}$. Consider
the continuum model $\Delta(k)=- k^2$. As shown in \cite{SM} it is natural to extend the
definition of the Larkin length to any $d,N$ as $L_c := (\frac{t_0^2}{12 B''(0)})^{\frac{1}{4-d}}$.
One then easily finds that for $\mu=0$ at large $N$, the mean number of equilibria grows exponentially
as $\overline{{\cal N}_{\rm tot} } \sim e^{C N (L/L_c)^d}$,
where $C$ is a numerical constant, given in \cite{SM}.

{\bf Complexity of local minima}.
To obtain the mean number of minima we consider
\bea \label{start2}
&& \overline{|\det {\cal K}[{\bf 0}]| \theta({\cal K}[{\bf 0}])} =  \prod_x \int_{\mathbb{R}}  \frac{d\xi(x) e^{- N \frac{\xi(x)^2}{2}}}{\sqrt{2 \pi/N}}\\
&& \times
\langle | \det ( K + X + \mu I )
\theta(K + X + \mu) | \rangle_{\rm GOE's} \nonumber
\eea
Let us define the domain ${\cal D}$ in the space of $\xi(x)$ as
\be
\lim_{N \to +\infty} \langle \theta(K + X + \mu) \rangle_{\rm GOE's} =1
\ee
{Using the ideas of Large Deviation Theory (again based on the spectral rigidity arguments)  \cite{footnote2} we conjecture that}
\bea
&& \langle | \det ( K + X + \mu I )
\theta(K + X + \mu) | \rangle_{\rm GOE's} \\
&& = \langle | \det ( K + X + \mu  | \rangle_{\rm GOE's} \quad , \quad \xi(x) \in {\cal D}
\eea
and is smaller by a factor $e^{- N^2 \Phi(X)}$ with $\Phi(X)>0$ when $\xi(x) \notin {\cal D}$.
Hence we can write
\be \label{start3}
\overline{|\det {\cal K}[{\bf 0}]| \theta({\cal K}[{\bf 0}]) }|_{N \gg 1} \sim \int_{\cal D} \prod_x
 \frac{d\xi(x)}{\sqrt{2 \pi/N}}  e^{- N S[\xi] }
\ee
with the same action as in \eqref{Sx}, but with an integration restricted to the domain ${\cal D}$.
Repeating the above minimization procedure, one finds that the uniform saddle point $\xi(x)=\xi_*$
belongs to ${\cal D}$ iff $\xi_* \geq - \lambda_e^- - \mu$ where $\lambda_e^-$ is the lowest
spectral edge of the support of the mean density. From Eq. \eqref{remark}
this condition is the same as $y(\mu)=0$ which defines the simple phase where
$\Sigma(\mu)=0$, hence it is equivalent to $\mu \geq \mu_c$.
Since the saddle point $\xi^*$ belongs to ${\cal D}$ in that case we
also get, not surprisingly, that $\Sigma_{st}(\mu)=0$.

In the complex phase $\mu<\mu_c$, the saddle point of $S[\xi]$ is outside the domain ${\cal D}$.
In this situation the integral will be dominated by the point of intersection
of the diagonal $\xi(x)=\xi$ with the boundary of the domain,
i.e. $\xi=\xi_e = - \lambda_e^- - \mu$. From Eqs. \eqref{solu2}
setting $y_\lambda \to 0$,
and using \eqref{refs}, one obtains the following expression for the lower spectral edge
\cite{footnote3}
\bea
\lambda^-_e = - \mu_c - \int_k \frac{1}{\mu_c - t \Delta(k)}
\eea
Substituting now $\xi_* \to \xi_e = - \lambda_e^- - \mu$ into \eqref{sig1}
gives after straightforward manipulations \cite{SM}

\be \label{SstA}
\Sigma_{\rm st}(\mu) =  - \frac{1}{2} ( \mu_c -  \mu)^2 + \int_{\mu}^{\mu_c}\,( I_1(\tilde \mu) - I_1(\mu_c)
 d \tilde \mu\,,
\ee
where the integrals $I_p(\mu)$ have been defined after eq.(\ref{expansion0}). This result gives for $d=0$, i.e. $L=1$,
$\Sigma_{\rm st}(\mu) =  - \frac{1}{2} [ (2  - \mu)^2 -1 ]  -  \ln  \mu$,
with $\mu_c=1$, in agreement with \cite{FyoWil07,FyoNad12}. It is worth noting that for the continuum model the
complexity of minima is given by UV convergent integrals for $d<4$, and takes an especially simple form \cite{SM}:
\be\label{mincompcontinuumA}
\Sigma_{st}(\delta)/\mu_c^2=-\frac{1}{2}\delta^2-\frac{2}{2-d}\delta
+\frac{4}{d(2-d)}\left[1-\left(1-\delta\right)^{d/2}\right],
\ee
where as before $\delta=1-\mu/\mu_c$. Note that for $d\to 2$ the right-hand side remains finite:
\[
\Sigma_{st}(\delta)/\mu_c^2|_{d=2}=-\frac{1}{2}\delta^2+\delta
+\left(1-\delta\right)\ln{\left(1-\delta\right)},
\]

Let us investigate the critical behavior of the complexity of stable equilibria.
Taking derivatives of \eqref{Sst} w.r.t. $\mu$ one finds that \cite{SM}:
\be\label{mincrit}
\Sigma'_{\rm st}( \mu_c) = \Sigma''_{\rm st}( \mu_c) = 0, \quad
\Sigma'''_{\rm st}(\mu_c) = - 2 I_3(\mu_c)
\ee
Hence, the third derivative is non-zero and the transition for
the complexity of the minima is of cubic order, i.e. $\Sigma_{\rm st}(\mu) \sim (\mu_c-\mu)^3$
for $\mu$ close to $\mu_c$.  For the continuum model $\Delta(k)=-k^2$ one
finds from \eqref{mincompcontinuumA} $\Sigma_{st}(\delta)/\mu_c^2 \simeq \frac{4-d}{12} \delta^3$ for
$\delta \ll 1$.
This corroborates with the third order nature of mean-field spin-glass phase transitions \cite{thirdSG},
and supports the view that the complexity of minima is thermodynamically relevant and is related to configurational entropy which dominates mean field spin-glass type transitions.

We now discuss the complexity in presence of an applied force $f(x)$ along the direction $i=1$
as described by Eq. \eqref{force1}.
If we consider the full space ${\bf u} \in \mathbb{R}^N$, the mean total number
of equilibria $\overline{{\cal N}_{\rm tot} }$ for $\mu>0$ is independent
of $f$, as is easily seen by performing a shift in ${\bf u}$.
Hence we calculate the mean number of equilibria $\overline{{\cal N}^w_{\rm tot} }$ in a
domain of effective finite width $w$ in all ${\bf u}$ directions, inserting the factor $\phi({\bf u})= e^{- {\bf u}^2/(2w^2)}$
in the r.h.s. of \eqref{ntot} (for more details see Section 4 in \cite{FLRTa}).
Following similar steps as in \cite{FLRTa} we obtain
\be \label{nw}
\overline{{\cal N}^w_{\rm tot} } =
\frac{ \overline{|\det {\cal K}[{\bf 0}]|}
e^{ - \frac{N}{2 w^2} \sum_{xy} f(x) [J^2 (\mu - t \Delta)^2 + \frac{2 B'(0)}{w^2}]^{-1}_{x,y} f(y)} }{[\det[(\mu - t \Delta)^2 + \frac{B'(0)}{2 B''(0) w^2}]^{N/2}}
\ee
a formula which is exact for any $N$, but which we study here for $N \to +\infty$.
The exponential factor reduces the number of equilibria. For a uniform
force $f(x)=f$, the value of $f$ at which the associated annealed
complexity attains zero is called $f_c^{\rm tot}$. The connection to
depinning is obtained in the limit $\mu \to 0$ performed before the
limit $w \to +\infty$ (see discussion in \cite{FLRTa}). In that limit
one finds from \eqref{nw}
\be
\overline{{\cal N}^w_{\rm tot} }|_{\mu \to 0, w\to 0} \sim e^{L^d N (\Sigma(\mu=0) - \frac{f^2}{4 B'(0)}) }
\ee
leading to
\be
f_c^{\rm tot} = \sqrt{ 4 B'(0) \Sigma(\mu=0) }
\ee
A similar
calculation can be performed for the stable equilibria. One finds
that the annealed complexity of minima in presence of a force vanishes
at
\be \label{fcst}
f_c^{\rm st} = \sqrt{ 4 B'(0) \Sigma_{\rm st}(\mu=0) } < f_c^{\rm tot}
\ee
where from \eqref{mincompcontinuumA}, $\Sigma_{\rm st}(\mu=0)=\mu_c^2\frac{4-d}{2d}$.

As discussed in the introduction this implies the bound \eqref{bound}
for the depinning threshold. It is sharper than the bound provided by
the total complexity
$f_c \leq f_c^{\rm tot}$ (a method used in the case $N=1$ in \cite{FLRTa}).
To know whether the bound \eqref{bound} obtained here provides the actual value of the
threshold would require estimating the
moments of the number of stationary points. Although work is in
progress in that direction, it is at present unclear if annealed and quenched complexities coincide
in this model.  In the pure $p$-spin model however it was shown to be the case \cite{Sub2017,RBBC2018},
while in other models such as spiked tensor \cite{RBBC2018},
and random field models \cite{TarjusRFIM}, it appears not
to be the case.

\bigskip

\acknowledgments

{\bf Acknowledgments:} We thank G. Biroli for discussions.
YVF thanks the Philippe Meyer Institute for Theoretical Physics
at ENS in Paris and the EPSRC grant  EP/N009436/1 "The many faces of random characteristic polynomials"
for support.
PLD acknowledges support from ANR grant ANR-17-CE30-0027-01 RaMa-TraF.
We also thank IIP at Natal for hospitality and support during the workshop "Random geometries
and multifractality in Condensed Matter and Statistical Mechanics".


{}

\newpage

.

\newpage

\begin{widetext}

\bigskip

\bigskip

\begin{large}
\begin{center}

{\bf SUPPLEMENTARY MATERIAL}

FOR

{\it Manifolds in high dimensional random landscape: complexity of stationary points
and depinning}

Yan V. Fyodorov$^1$ and Pierre Le Doussal$^2$

\end{center}
\end{large}

$^1$ King's College London, Department of Mathematics, London  WC2R 2LS, United Kingdom.

$^2$ Laboratoire de Physique Th\'eorique de l'Ecole Normale Sup\'erieure,
PSL Research University, CNRS, Sorbonne Universit\'es,
24 rue Lhomond, 75231 Paris, France.

\bigskip

We give the principal details of the calculations described in the manuscript of the Letter.

\section{1. Expansion near the transition}

Taking derivatives of Eq. \eqref{SS} in the Letter we obtain
\be
\Sigma''(\mu) = \partial_\mu \xi_* + I_2(\mu)
\ee
Let us define for convenience
\be\label{defIp}
I_{pq}(\mu,y^2) := \int_k \frac{(\mu - t \Delta(k))^q}{((\mu - t \Delta(k))^2 + y^2)^{p/2}}
\quad , \quad I_p(\mu) := \int_k \frac{1}{(\mu - t \Delta(k))^p}
\ee
The equations \eqref{solu} of the text which determine $y$ and $\xi^*$ read for $\mu<\mu_c$
\bea
&& \xi_* = I_{21}(\mu,y^2) \\
&& 1 = I_{20}(\mu,y^2)
\eea
We will need the following relations which easily follow from the definition (from now on we suppress the arguments $\mu,y^2$ of all $I_{pq}$ integrals)
\bea
&& \partial_\mu I_{20}=- 2 I_{41} \quad , \quad \partial_{y^2} I_{20}=-  I_{40} \\
&& \partial_\mu I_{21}=I_{20}- 2 I_{42} \quad , \quad \partial_{y^2} I_{21}=-  I_{41}
\eea

Taking a derivative of the second SP equation we obtain
\be
\frac{d y^2}{d \mu} = - \frac{\partial_\mu I_{20}}{\partial_{y^2} I_{20}} =
-2 \frac{I_{41}}{I_{40}}
\ee
The derivative of the first SP equation gives
\be
\partial_\mu \xi_* = \partial_\mu I_{21} + \frac{d y^2}{d \mu} \partial_{y^2} I_{21}
= I_{20}- 2 I_{42} + 2 \frac{I_{41}^2}{I_{40}}
\ee
which is exact for arbitrary $\mu<\mu_c$. Taking the limit
$\mu \to \mu_c^-$ we obtain, since $y(\mu_c)=0$
\be
\partial_\mu \xi_* = - I_2(\mu_c) + 2 \frac{I_3(\mu_c)^2}{I_4(\mu_c)}
\ee
Using that $I_2(\mu_c)=1$ we obtain
\be
\Sigma''(\mu_c)= 2 \frac{I_3(\mu_c)^2}{I_4(\mu_c)}
\ee
As $\Sigma(\mu_c)=\Sigma'(\mu_c)=0$ this immediately implies (\ref{expansion0}).\\

\noindent {\bf Complexity of minima for $\mu\le \mu_c$.}

Substituting $\xi_* \to \xi_e = - \lambda_e^- - \mu$ into \eqref{sig1} we get
\bea \label{Sst}
&& \Sigma_{\rm st}(\mu) =  - \frac{1}{2} [ ( \mu_c + \int_k \frac{1}{ \mu_c - t \Delta(k)}  -  \mu)^2
\\
&& - ( \int_k \frac{1}{ \mu_c - t \Delta(k)})^2]  -  \int_k [ \ln (\mu - t \Delta(k)) -  \ln(\mu_c - t \Delta(k)) ] \nonumber
\eea
where we have subtracted the value $\Sigma_{\rm st}(\mu_c)=0$ which allows
to eliminate the constant $f(- \lambda_e^-)$. Finally, by differentiating over the parameter $\mu$ it is
easy to show that
\[
\int_k [ \ln (\mu - t \Delta(k)) -  \ln(\mu_c - t \Delta(k)) ]=-\int_{\mu}^{\mu_c}I_1(\tilde{\mu})d\tilde{\mu}.
\]
 Expanding the square and reordering we obtain the formula
(\eqref{SstA}) in the text.

Let us investigate the critical behavior of the complexity of stable equilibria.
Taking derivatives of \eqref{SstA} w.r.t. $\mu$ we find, using the definition \eqref{refs}
of $\mu_c$ in the second line we get
\bea
&& \Sigma'_{\rm st}( \mu) = - (\mu - \mu_c - I_1(\mu_c)) - I_1(\mu) \Rightarrow
\Sigma'_{\rm st}( \mu_c) = 0 \nonumber \\
&& \Sigma''_{\rm st}( \mu) = - 1 +  I_2(\mu) \Rightarrow
\Sigma''_{\rm st}( \mu_c) = 0 \\
&& \Sigma'''_{\rm st}(\mu) = - 2 I_3(\mu) \Rightarrow
\Sigma'''_{\rm st}(\mu_c) = - 2 I_3(\mu_c) \nonumber
\eea
\section{2. Explicit formulas for the complexity in the continuum model}

\noindent{\bf Total complexity.} Here we analyze the equations \eqref{SS2} and \eqref{ymu} which
determine $\Sigma(\mu)$ as a function of $\mu$ in the complex phase $\mu<\mu_c$.

Let us consider the continuum model in dimension $d$, with  $\Delta(k)=- k^2$. We restrict to
$d<4$.
We assume (and check later) that the momentum integrals are convergent for $k \in \mathbb{R}^d$.
Upon scaling $k=\sqrt{\tilde \mu/t} p$ and $y=\tilde \mu x$  and  employing spherical coordinates it turns out to be useful to introduce the following functions:
\be \label{deff}
 f_d(x)=C_d \int_{0}^{\infty} \frac{dq \, q^{\frac{d-2}{2}}}{(1+q)^2 + x^2}
 \quad , \quad g_d(x)=C_d x^2 \int_{0}^{\infty} \frac{dq \, q^{\frac{d-2}{2}}}{[(1+q)^2 + x^2](1+q)}
\ee
where $C_d= \frac{S_d}{2 (2 \pi)^d}  = \frac{1}{2^d\pi^{d/2}\Gamma(d/2)}$, with $S_d$ standing for the
area of the hypersphere in dimension $d$. With help of the introduced notations (\ref{deff}) the equation (\ref{ymu}) takes the form
\be \label{complexitycont1}
1 = t^{-d/2}\mu^{\frac{d-4}{2}} f_d\left(\frac{y}{\mu}\right)
\ee
which, in particular, implies that the Larkin mass satisfies the relation $t^{d/2}\mu_c^{\frac{4-d}{2}} = f_d(0)$.
Solving (\ref{complexitycont1}) by functional inverse as $y(\mu)=\mu\,f_d^{-1}\left(t^{d/2}\mu^{\frac{4-d}{2}}\right)$
allows us to write the complexity (\ref{SS2}) explicitly  as
\be \label{complexitycont2}
\Sigma(\mu) = t^{-d/2}\int_{\mu}^{\mu_c} \frac{d\tilde \mu}{\tilde \mu^{\frac{2-d}{2}}} g_d(f_d^{-1}(\tilde \mu^{\frac{4-d}{2}}t^{d/2}))
\ee
Further changing $\tilde \mu=\left(xt^{-d/2}\right)^{\frac{2}{4-d}}$ and $\mu=\mu_c(1-\delta)$ the above can be presented in the form
\be \label{complexitycont3}
\Sigma(\delta) =  \frac{2}{4-d}  t^{-2d/(4-d)}
 \int^{f_d(0)}_{f_d(0)(1-\delta)^{\frac{4-d}{2}}} dx{x^{\frac{2(d-2)}{4-d}}} g_d\left[f_d^{-1}(x)\right]
\ee
implying, in particular
\be \label{complexitycont3a}
 \frac{d}{d\delta}\Sigma(\delta) = t^{-2d/(4-d)} f_d(0)^{\frac{d}{4-d}}(1-\delta)^{\frac{d-2}{2}}
g_d\left[f_d^{-1}\left(f_d(0)(1-\delta)^{\frac{4-d}{2}}\right)\right]
\ee
 On the other hand, setting $\delta\to 1$ is equivalent to $\mu\to 0$. The zero mass limit can be easily
found from (\ref{complexitycont3}). After substituting  $z= f^{-1}_d(x)$ and
taking into account $f_d(x\to \infty)=0$ one gets
\be \label{zeromd}
\Sigma(\mu=0) = \sigma_d t^{-2d/(4-d)} \quad , \quad
\sigma_d = \frac{2}{4-d}
 \int_{0}^{+\infty} dz |f'_d(z)| {f_d(z)^{\frac{2(d-2)}{4-d}}} g_d(z)
\ee

{\bf Behavior near $\mu_c$}. Let us define, for $p > d/2$
\be \label{Itilde}
\tilde I_p := C_d\,\int_{0}^{\infty} \frac{q^{\frac{d}{2}-1}\, dq}{(1+q)^p}
= \frac{1}{2^{d} \pi ^{d/2}} \frac{ \Gamma
   \left(p-\frac{d}{2}\right)}{\Gamma (p)}\,.
 \ee
Then expanding in (\ref{deff}) as
\[
f_d(x\ll 1)=f_d(0)-x^2\tilde{I}_4 +o(x^2), \quad g_d(x\ll 1)=x^2 \tilde{I}_3+o(x^2) \quad , \quad f_d(0)=\tilde I_2
\]
it is easy to deduce from (\ref{complexitycont1}) that to the leading order in $\delta\ll 1$
holds
\[
f_d^{-1}\left(f_d(0)(1-\delta)^{\frac{4-d}{2}}\right)\simeq A\, \delta^{1/2},
\]
where the coefficient $A$ is given by
\be \label{A}
A^2=\frac{f_d(0)(4-d)}{2\tilde I_4},
\ee
Substituting all those expressions to (\ref{complexitycont3}) gives to the leading order
\be \label{complexitycontleading1}
\frac{d}{d\delta}\Sigma(\delta) \simeq  t^{-2d/(4-d)} f_d(0)^{\frac{d}{4-d}}A^2\tilde{I}_3\delta
= \frac{4-d}{2}t^{-2d/(4-d)} f_d(0)^{\frac{4}{4-d}}\frac{\tilde{I}_3}{\tilde{I}_4}\, \delta\,
\ee
and further using $\mu_c = \left(f_d(0)/t^{d/2}\right)^{\frac{2}{4-d}}$ and $\frac{\tilde{I}_3}{\tilde{I}_4}=\frac{6}{6-d}$ we see that
\[
\frac{d}{d\delta}\Sigma(\delta\ll 1) \approx 3\frac{4-d}{6-d}\mu_c^2 \, \delta
\]
As $\Sigma(\delta=0)=0$ this finally implies that for the continuum model with $\Delta(k)=-k^2$ the complexity close to the threshold is given by
\be\label{expansion0cont}
\Sigma(\delta\ll 1) \approx \frac{3}{2}\,\frac{4-d}{6-d}\mu_c^2 \, \delta^2
\ee
This fully agrees with the general expression (\ref{expansion0}). Indeed, it is easy to see that in the continuum limit
the integrals $I_p$ defined in (\ref{defIp}) are related to $\tilde{I}_p$ as
\be \label{Ipcont}
I_p(\mu_c)=\tilde{I}_p\mu_c^{\frac{d}{2}-p} t^{-d/2}
\ee
so that again using $\mu_c^{\frac{4-d}{2}}\,t^{d/2} = \tilde{I}_2$ we see
\[
\frac{I_3^2(\mu_c)}{I_4(\mu_c)}=\frac{\tilde I_3^2}{\tilde I_4} \mu_c^{\frac{d}{2}-2}t^{-d/2}= \frac{\tilde I_3^2}{\tilde I_4\tilde I_2}
=\frac{3}{2}\,\frac{4-d}{6-d}
\]
exactly as expected. \\

\noindent{\bf Complexity of minima.} Let us evaluate the
complexity of minima \eqref{SstA} for the continuum model, $\Delta(k)=-k^2$.
Although each factor $I_1$ is a divergent integral (and would require a UV cutoff)
for $d \geq 2$, the difference
\be
I_1(\tilde \mu) - I_1(\mu_c) = \int_{\tilde \mu}^{\mu_c} d\rho \, I_2(\rho)
= t^{-d/2} \int_{\tilde \mu}^{\mu_c} d\rho \, \rho^{\frac{d}{2}-2}
= t^{-d/2}  \frac{2 \tilde I_2}{d-2} ( \mu_c^{\frac{d-2}{2}} - \tilde \mu^{\frac{d-2}{2}} )
\ee
is convergent for any $\tilde \mu \geq 0$ for $d<4$. Inserting into \eqref{SstA},
and remembering that $t^{-d/2} \mu_c^{\frac{d}{2}-2} \tilde I_2=1$,
we obtain upon integrating once more, the complexity of minima (\eqref{SstA}) in the form
\be\label{mincompcontinuum}
\Sigma_{st}(\mu<\mu_c)/\mu_c^2=-\frac{1}{2}\left(1-\frac{\mu}{\mu_c}\right)^2-\frac{2}{2-d}\left(1-\frac{\mu}{\mu_c}\right)
+\frac{4}{d(2-d)}\left[1-\left(\frac{\mu}{\mu_c}\right)^{d/2}\right]
\ee
which upon substituting $\mu/\mu_c=1-\delta$ gives (\eqref{mincompcontinuumA}) in the text.
 This expression is valid for $d<4$ and has a finite limit for $d=2$ as given in the text. \\

\noindent {\bf Results in two dimensions, $d=2$}.
In this case the functions $f_2(x)$ and $g_2(x)$ can be found explicitly as:
\bea
&& f_2(x) = \frac{1}{4 \pi x} \left( \frac{\pi}{2} - \tan ^{-1}( \frac{1}{x} ) \right) = \frac{\tan ^{-1} x}{4 \pi x}
=
\frac{1}{4 \pi }-\frac{x^2}{12 \pi }+\frac{x^4}{20 \pi
   }+O\left(x^5\right) \\
&& g_2(x) = \frac{\log(1+x^2)}{8 \pi}
\eea
 We now use Eq. \eqref{complexitycont2} which reads in $d=2$
\be\label{intcomp}
\Sigma(\mu) = \frac{1}{t} \int_{\mu}^{\mu_c} d\tilde \mu \, g_2(f_2^{-1}(\tilde \mu t))
= - \frac{1}{t^2} \int_{0}^{Z(\mu)} dz f'_2(z)  \, g_2(z)
\ee
where we have introduced $z$ as the solution to $t \tilde \mu=f_2(z)$,
$Z=Z(\mu)$ as the solution to $t  \mu=f_2(Z)$, and used that $t \mu_c = f_2(0)=\frac{1}{4 \pi}$.
 The integral (\ref{intcomp}) can be easily evaluated by parts yielding
for the complexity an explicit parametric system where $Z$ must be eliminated
\bea
&& \Sigma   = - \frac{\tan ^{-1}(Z) \left(\log \left(Z^2+1\right)-Z
   \tan ^{-1}(Z)\right)}{32 \pi ^2 t^2 Z} \\
   && \mu=t^{-1}f_2(Z) = \frac{\tan ^{-1} Z}{4 \pi t Z}
\eea
which can be further written as
\bea
\Sigma =  \frac{\mu}{8 \pi t} \left(4 \pi t \mu Z^2 -  \log(1 + Z^2)  \right) \quad , \quad
\mu=\frac{\tan ^{-1} Z}{4 \pi t Z}  \quad , \quad \mu_c= \frac{1}{4 \pi t}
\eea
or equivalently
\bea
\frac{\Sigma}{\mu_c^2} =  \frac{\mu}{2 \mu_c} \left(\frac{\mu}{\mu_c} Z^2 -  \log(1 + Z^2)  \right) \quad , \quad
\frac{\mu}{\mu_c}=\frac{\tan ^{-1} Z}{Z}
\eea

In particular, we obtain the series expansion
close to the transition when $\mu = \mu_c (1- \delta)$ with $\delta \ll 1$
\bea
\Sigma(\delta\ll 1) = \mu_c^2\left[\frac{3 \delta ^2}{4}+\frac{3 \delta ^3}{20}+\frac{117
   \delta ^4}{1400}+\frac{351 \delta
   ^5}{7000}+O\left(\delta ^6\right)\right] \quad ,
\eea
where the first term agrees with the general result (\ref{expansion0cont}).

 The limit $\mu \to 0$ corresponds to $Z \to +\infty$ and we obtain the small $\mu$ expansion
\be
\frac{\Sigma}{\mu_c^2} = \frac{\pi ^2}{8}+\frac{\mu}{\mu_c} \left(\log \left(
   \frac{2 \mu}{\pi \mu_c} \right)-1\right)+\frac{2 \mu^2}{\pi ^2 \mu_c^2}-\frac{2 \left(\pi ^2-12\right) \mu^3}{3 \pi ^4 \mu_c^3}+O\left((\frac{\mu}{\mu_c})^4\right) \quad , \quad \mu_c= \frac{1}{4 \pi t}
\ee
In particular we find the the finite value in $d=2$
\be
\Sigma(\mu=0)|_{d=2}=\frac{1}{128 t^2}
\ee

{\bf Results in dimension one,  $d=1$}
 For the continuum model in $d=1$ the complexity \eqref{complexitycont3}
can be calculated inserting
\bea
&& f_1(x) = \frac{i}{4 x} ( \frac{1}{\sqrt{1+ i x}} - \frac{1}{\sqrt{1- i x}} ) \\
&& g_1(x) = - \frac{1}{4} ( \frac{1}{\sqrt{1+ i x}} + \frac{1}{\sqrt{1- i x}} -2 )
\eea
 Since we did not find a simpler expression in $d=1$ we give here only a
numerical evaluation for zero mass, from \eqref{zeromd}
\be
\Sigma(\mu=0)|_{d=1} \approx 0.375 \,  t^{-2/3}
\ee

\section{3. Larkin length}

There are several conventions to define the Larkin length $L_c$, and they simply differ by some constant prefactors in
the weak disorder limit. Let us consider here the continuum model $\Delta(k)=-k^2$.

If we stick to the definition $L_c = (\kappa^2/R''''(0))^{1/3}$ given for $N=1$, $d=1$ in \cite{FLRT}, the correspondence is that
$\kappa$ there equals $t_0$ here, and $R(u)$ there is $R(u)=B(u^2)$,  which, in particular, gives the relation between the derivatives: $R''''(u)=12B''(u^2)+48u^2B'''(u^2)+16u^4B''''(u^2)$, hence $R''''(0)=12B''(0)$. To remain consistent with
 the convention in \cite{FLRT}, we then define for the case of general $N,d$
\be \label{defLc2}
L_c := \left(\frac{t_0^2}{12 B''(0)}\right)^{\frac{1}{4-d}} = (t^2/3)^{\frac{1}{4-d}}
\ee
where we recalled that $t=t_0/2\sqrt{B''(0)}$.

In general we expect, for the complexity defined in the large $L$ limit
\be \label{s11}
\Sigma(\mu=0) = C_{N,d} L_c^{-d}
\ee
where $C_{N,d}$ is a constant prefactor.  In \cite{FLRT} it was numerically found that
$C_{1,1}\approx 0.46$. Here we show that in the large $N$ limit \eqref{s11}
indeed holds with
\be
\lim_{N \to +\infty} C_{N,d}  = C_{\infty,d} = \sigma_d 3^{- \frac{d}{4-d}}
\ee
where the last equality is obtained by comparing \eqref{defLc2}, \eqref{s11} and the result \eqref{zeromd}  for $\Sigma(\mu=0)$
where the constant $\sigma_d$ was defined. We thus obtain, for different dimensions:
\be
C_{\infty,d=1} = 0.260.. \quad , \quad C_{\infty,d=2} = 0.00260
\ee \\

{\bf Universal ratio.}
Finally it is interesting to consider the dimensionless ratio $\frac{\Sigma_{st}(\mu)}{\Sigma(\mu)}$ for
$\mu<\mu_c$.
It vanishes linearly near $\mu=\mu_c$, whereas at $\mu=0$, using the relation $\mu_c=(\tilde I_2 t^{-d/2})^{\frac{2}{4-d}}$, its value for the continuum model is a universal number (in $[0,1]$) depending only on $d$:
\be
\frac{\Sigma_{st}(\mu=0)}{\Sigma(\mu=0)} = \frac{\frac{4-d}{2d} \mu_c^2}{ \sigma_d t^{-2 d/(4-d)}}
= \frac{4-d}{2d \sigma_d} (\tilde I_2)^{\frac{4}{4-d}}
\ee
where $\sigma_d$ is defined in \eqref{zeromd}. This number is $0.63..$ for $d=1$ and $0.405..$ for $d=2$.

\section{Limit $q\to 0$ and resolvant}

Here we check that the real part of the mean resolvant of the Hessian is correctly predicted
by our theory. Let us consider the Hessian ${\cal K}^0[{\bf u}] \to {\cal K}[{\bf u}]$ around
configuration ${\bf u}$ defined in the text in \eqref{Hessian0} (we recall that in our units
$\mu_0 \to \mu$, $t_0 \to t$ and $J^2=4 B''(0) \to 1$). It was also defined in
Eq (3) in \cite{UsHess}, together with the Green function (Eqs. (8,9,23) there)
\be
{\cal G}(x,y;\lambda,{\bf u}) = \frac{1}{N} \sum_{i=1}^N \left(\frac{1}{\lambda - {\cal K}[{\bf u}]}\right)_{xi,yi}
\ee
and with the mean resolvant
\be
{\cal G}(\lambda,{\bf u}) = \frac{1}{L^d} \sum_x \overline{ {\cal G}(x,x;\lambda,{\bf u}) }
\ee
where here ${\bf u}$ is a fixed typical configuration, which can be chosen to be ${\bf u}={\bf 0}$.
In that case the block matrix covariance structure is recalled in (6-7) there and is related to the one
of Wegner orbital models, and in the continuum limit in $x$, to matrix Anderson models.
The mean resolvant was calculated in \cite{UsHess} (in agreement with earlier results by Pastur)
with the result that it is the solution of the self-consistent equation (for $i p$)
\be \label{pastur}
{\cal G}(\lambda,{\bf 0}) = i p  \quad , \quad i p = \int_k \frac{1}{\lambda - \mu + t \Delta(k) - i p}
\ee

On the other hand, the following equality allows an independent calculation of the mean resolvant
\be \label{der0}
\frac{1}{N L^d} \partial_q|_{q=0} \partial_\mu \overline{|\det {\cal K}[{\bf 0}]|^q}
= \frac{1}{N L^d} \partial_\mu \overline{ \log |\det {\cal K}[{\bf 0}]| } = \frac{1}{N L^d} \partial_\mu
 {\rm Re} \overline{ {\rm Tr} \log {\cal K}[{\bf 0}] }
=  \frac{1}{N L^d}  {\rm Re} \overline{ {\rm Tr} {\cal K}[{\bf 0}]^{-1} } = -  {\rm Re} {\cal G}(\lambda=0,{\bf 0})
\ee
Now we can evaluate $\overline{|\det {\cal K}[{\bf 0}]|^q}$, for any $q$, by a simple generalization of
the calculation in this paper, which corresponds to the particular case $q=1$. The case of general $q$ will be detailed and analyzed in a forthcoming publication \cite{inprep}, here we just sketch the calculation in the limit $q \to 0$. It is easy to see, that under the same assumptions as for $q=1$
\be \label{start2n}
\overline{|\det {\cal K}[{\bf 0}]|^q}|_{N \gg 1} \sim \prod_x \int_{\mathbb{R}}
 \frac{d\xi(x)}{\sqrt{2 \pi/N}}  e^{- N S[\xi] }
\quad , \quad
S[\xi]= \sum_x \frac{1}{2} \xi(x)^2 - \frac{q}{N} \left\langle {\rm Tr}  \log | K + X +  \mu I |  \right\rangle_{\rm GOE's}
\ee
The natural saddle point at large $N$ is $\xi(x)=\xi^*_q$, where $\xi^*_q$ solves the
equation
\be \label{sp0n}
\xi^*_q = q f'(\xi^*_q + \mu)  \quad , \quad  f(\xi) := \int d\lambda \ln|\lambda+\xi| \, \rho_{K}(\lambda)
\ee
From which we obtain
\be
\frac{1}{N L^d} \partial_\mu \log \overline{|\det {\cal K}[{\bf 0}]|^q} = \partial_\mu \left( - \frac{1}{2} (\xi_q^*)^2 +  q f(\xi_q^* + \mu) \right) = q f'(\xi^*_q + \mu) = \xi^*_q
\ee
Hence, taking a derivative w.r.t. $q$ and using \eqref{der0} we obtain that by this method the
mean resolvant is obtained as
\be \label{resres}
 {\rm Re} \, {\cal G}(\lambda=0,{\bf 0}) = - \partial_q|_{q=0} \xi^*_q := - G
\ee
where $G$ is by definition the leading order
in the Taylor expansion of $\xi^*_q = q G + O(q^2)$ at small $q$.

Now it is easy to see that \eqref{sp2} in the text generalizes into
\bea \label{sp2n}
\xi_q^* = - q {\rm Re} [ i r_{- \xi_q^*-  \mu + i 0^+} ]
\eea
which, in the limit $q \to 0$ becomes
\bea \label{22}
G = - {\rm Re} [ i r_{-  \mu + i 0^+} ]
\eea
From \eqref{sc} in the text we see that $i r_{- \mu}$ satisfies
\be \label{scn}
i r_{-\mu} = \int_k \frac{1}{- \mu + t \Delta(k)   - i r_{-\mu}}
\ee
Comparing with \eqref{pastur} we see that $i r_{-\mu}= i p$,
hence \eqref{22} implies $G = - {\rm Re} (i p)$, hence
\eqref{resres} leads to the correct result for the real
part of the mean resolvant ${\rm Re} \, {\cal G}(\lambda=0,{\bf 0})$.

{}

\bigskip

\end{widetext}


\begin{thebibliography}{}
\bibitem{footnote4}
For $N>1$ the Middleton theorems do not hold, so it cannot be
excluded that the manifold may get pinned or not depending on the initial condition.
\bibitem{footnote3} The expression for $\lambda_e^-$ was analyzed
in Ref. \cite{UsHess} for a more general problem which involved a parameter $\mu_{\rm eff}$
and is given there in (60).
The correspondence between the two problems is obtained by
setting $\mu_{\rm eff}=0$ as can be seen by comparing \eqref{sc}
with the self-consistent equation (52) identifying $p$ there with
$r_\lambda$ here.
\end{thebibliography}
\end{document}